\documentclass{article}
\usepackage{spconf}
\usepackage{graphicx}

\usepackage{tikz}
\usepackage{comment}
\usepackage{amsmath,amssymb} 
\usepackage{color}
\usepackage[utf8]{inputenc} 
\usepackage[T1]{fontenc}    
\usepackage{url}            
\usepackage{booktabs}       
\usepackage{amsfonts}       
\usepackage{nicefrac}       
\usepackage{microtype}      
\usepackage{xcolor}         

\usepackage[normalem]{ulem}
\usepackage{graphicx}
\usepackage{subfigure}
\usepackage{url}
\usepackage{multicol}
\usepackage{multirow}
\usepackage{wrapfig}
\usepackage{float}
\usepackage{tabularx}
\usepackage{enumitem}
\usepackage{soul}
\usepackage{array}

\usepackage{hyperref}
\usepackage{algorithmic}
\usepackage{algorithm}

\usepackage{amsthm}

\usepackage[capitalize]{cleveref}
\crefname{section}{Sec.}{Secs.}
\Crefname{section}{Section}{Sections}
\Crefname{table}{Table}{Tables}
\crefname{table}{Tab.}{Tabs.}
\usepackage{dsfont}
\usepackage{wrapfig}

\iftrue
\newcommand{\berivan}[1]{{\color{red}\textbf{BI}: #1}}
\newcommand{\phil}[1]{{\color{blue}\textbf{PAC}: #1}}
\fi

\usepackage{epsfig}
\usepackage{amsmath}
\usepackage{amssymb}
\usepackage{color}
\usepackage{url}
\usepackage{subfigure}

 \title{Sandwiched Video Compression: Efficiently Extending the Reach of Standard Codecs with Neural Wrappers}
 \name{Berivan Isik$^{\ast}$ \thanks{Work done while the first author was an intern at Google.}, 
 Onur G. Guleryuz$^{\dag}$, Danhang Tang$^{\dag}$, Jonathan Taylor$^{\dag}$, and Philip A. Chou$^{\dag}$
       }
 \address{$^{\ast}$Stanford University, $^{\dag}$Google Research \\
          \url{berivan.isik@stanford.edu}, \\
          \url{{oguleryuz, danhangtang, jontaylor, philchou }@google.com}}

\begin{document}
 \ninept
 \maketitle
\begin{abstract}
We propose sandwiched video compression -- a video compression system that wraps neural networks around a standard video codec. The sandwich framework consists of a neural pre- and post-processor with a standard video codec between them. The networks are trained jointly to optimize a rate-distortion loss function with the goal of significantly improving over the standard codec in various compression scenarios. End-to-end training in this setting requires a differentiable proxy for the standard video codec, which incorporates temporal processing with motion compensation, inter/intra mode decisions, and in-loop filtering. 
We propose differentiable approximations to key video codec components and demonstrate that, in addition to providing meaningful compression improvements over the standard codec, the neural codes of the sandwich lead to significantly better rate-distortion performance in two important scenarios.
When transporting high-resolution video via low-resolution HEVC, the sandwich system obtains 6.5 dB improvements over standard HEVC. More importantly, using the well-known perceptual similarity metric, LPIPS, we observe $~30 \%$ improvements in rate at the same quality over HEVC. Last but not least, we show that pre- and post-processors formed by very modestly-parameterized, light-weight networks can closely approximate these results.\footnote{Published at the International Conference on Image Processing (ICIP), 2023.}
\end{abstract}
\section{Introduction}
\label{sec:introduction}
The last decade 
has witnessed neural image compression methods \cite{balle2016end_pcs} gradually outperforming traditional image compressors such as JPEG2000 and HEVC \cite{sullivan2012overview} and replacing them as state-of-the-art image compressors. The success of neural image compressors is due to their ability to perform complex non-linear transforms, which are learned end-to-end through back-propagation \cite{balle2020nonlinear}. More recently, the first learned video compressors have been proposed \cite{rippel2019learned} and research on improving neural video compressors has accelerated \cite{agustsson2020scale, hu2021fvc, rippel2021elf, yang2020hierarchical}. The existing strategies focus mainly on learning how to predict flows, 
warp the previous reconstructions with the predicted flows,
and do residual compensation, all with neural networks trained end-to-end. 
This has lead to very complex networks with tens of millions of parameters \cite{hu2021fvc, rippel2021elf}, tens of millions of floating point operations per pixel, and weeks of training on high end GPUs \cite{lu2019dvc}.

\begin{figure}[t]
    \centering
    \includegraphics[width=.95\linewidth]{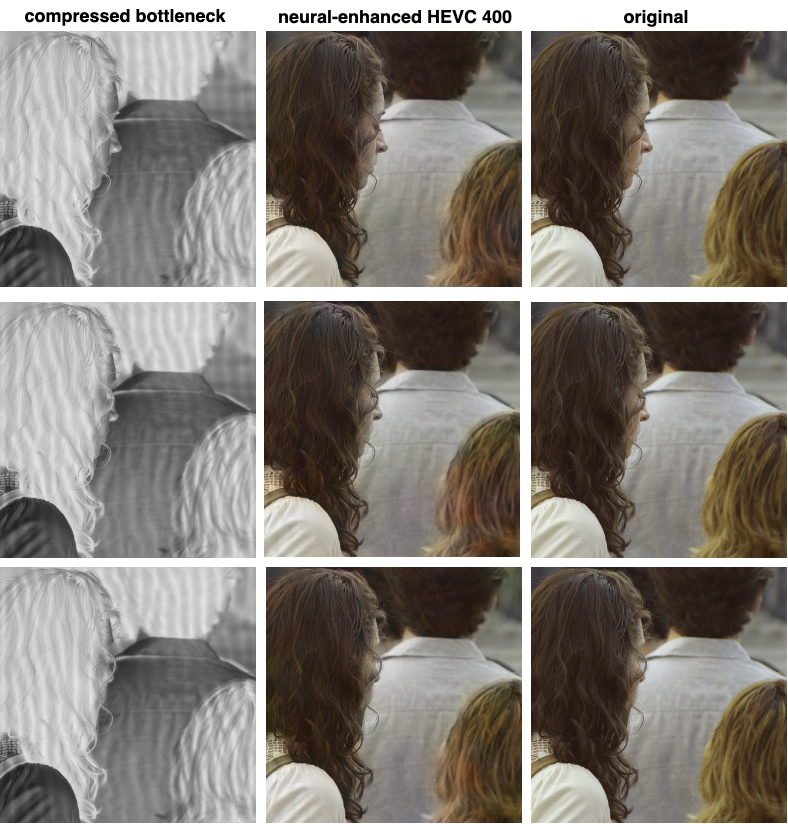}
    \caption{Toy example with gray-scale codec (HEVC4:0:0). The sandwich is used to transport full color video over a codec that can only transport gray-scale. Frames of neural codes decoded by the standard codec (compressed bottlenecks), final reconstructions by the post-processor, and original source videos, at time $t$, $t+1$, and $t+2$ are shown. Rate=0.50 bpp, PSNR=39.3dB, fps=30. Note that the sandwich establishes temporally consistent modulation-like patterns on the bottlenecks through which the pre-processor encodes color that is then demodulated by the post-processor for a full-color result. }
    \label{fig:subjective_YUV}
\end{figure}

In this work, we propose a computationally more efficient way of improving upon any standard video codec by using neural networks.
In particular, we sandwich a standard video codec between lightweight neural pre- and post-processors, and optimize them end-to-end.
Our approach leverages existing highly optimized computational implementations of the standard codec and the seamless network-transport of standard bit-streams over network nodes already familiar with such traffic.  The proposed neural pre- and post-processors learn how to perform non-linear transformations on the input and the output of the standard codec for the best rate-distortion performance while the standard video codec does the heavy lifting by performing the usual video coding operations on the output of the neural pre-processor. Our work is primarily geared toward adapting the standard codec to compression scenarios  that are outside the immediate scope of its design. In doing so, we are interested not only in significant improvements but also in
understanding how the neural networks accomplish those improvements.
As the networks have to generate viewable 8-bit video that is related to the original visual scenes, it is typically possible to observe their operation even if at a coarse level (see for instance the toy example of Figure \ref{fig:subjective_YUV}).

\begin{figure*}[t]
    \centering
    \includegraphics[width=\linewidth]{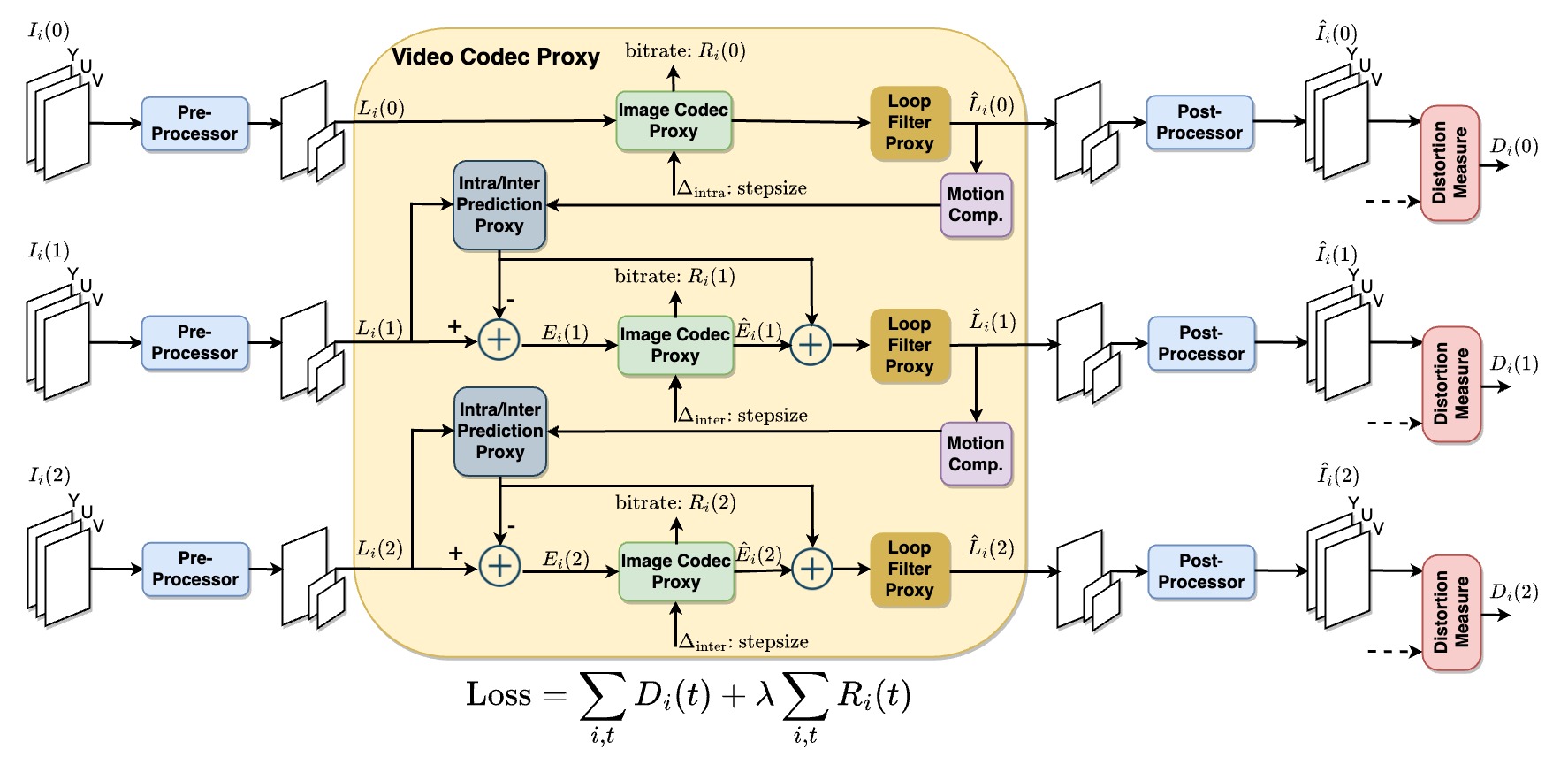}
    \caption{Sandwich architecture during training. The shaded box (video codec proxy) is replaced with a standard video codec at inference.}
    \label{fig:diagram}
\end{figure*}

Figure~\ref{fig:diagram} summarizes our approach.
For the $i$th video clip with frames $t=0,1,2,\ldots$, source images $\{I_i(t)\}$ are  processed by a neural pre-processor in a temporally independent fashion to obtain ``bottleneck'' images $\{L_i(t)\}$.  The bottleneck images contain neural latent codes representing the source images, but are nevertheless in 4:4:4 (or 4:2:0) format ready for compression by the standard video codec.  The standard video codec compresses the bottleneck images into reconstructed bottleneck images $\{\hat L_i(t)\}$, again in 4:4:4 format.  Finally, the reconstructed bottleneck images are processed by a neural post-processor into reconstructed source images $\{\hat I_i(t)\}$.  To train the neural pre- and post-processors, we replace the standard video codec by a differentiable proxy (shown shaded in yellow) and we minimize the total rate-distortion Lagrangian $\sum_{i,t}D_i(t)+\lambda R_i(t)$, where $D_i(t)$ and $R_i(t)$ are the distortion and rate of frame $t$ in clip $i$.  At evaluation, the standard video codec is used instead of the proxy.

 Our framework is inspired by previous work on sandwiched image compression \cite{guleryuz2021sandwiched, guleryuz2022sandwiched}, which shares the motivation to improve over standard codecs in rate-distortion performance, while leveraging their implementation-related conveniences. The end-to-end learned image codecs that outperform the standard image codecs in rate-distortion performance \cite{balle2016end_pcs, balle2018variational, hu2021learning, toderici2017full} require over-parameterized networks and entropy models, which have significant time, power, and resource costs. \cite{guleryuz2021sandwiched,guleryuz2022sandwiched} proposed a solution to this by wrapping 
 neural networks around a standard image codec. Training the sandwich end-to-end requires a differentiable proxy for the standard image codec. 
 The authors trained the sandwiched model by replacing the standard image codec with a differentiable JPEG proxy that uses $8\times8$ discrete cosine transforms (DCTs) and straight-through quantization\footnote{A straight-through quantizer is a software-implemented function $y=Q(x)$ that implements ordinary quantization in the forward direction but pretends to have a gradient $dy/dx$ of 1 instead of 0 in the backward direction.} as a proxy for the quantization step, and a differentiable function as an approximation of the bit rate from the entropy coding step. 
 They showed that with training, the pre-processor and post-processor would learn to communicate with each other by sending neural codes in the form of spatial modulation patterns that represent details of the original input. The patterns are easy to compress with a standard codec while being robust to its degradations.  Gains over the standard codec alone were shown to be especially noteworthy when the input has different characteristics than intended for the standard codec.  In particular, \cite{guleryuz2021sandwiched} showed high gains when coding multispectral or normal map images, while \cite{guleryuz2022sandwiched} showed high gains when coding high dynamic range or high resolution images with a low dynamic range (LDR) or low resolution (LR) codec.
 
 Our motivation and approach are similar to \cite{guleryuz2021sandwiched,guleryuz2022sandwiched} but in this paper, we make the fundamental leap of transferring the sandwich approach from images to video. This leap is challenging for a number of reasons.  First, independently pre- and post-processing the video frames may be problematic, as the neural codes in the bottleneck images need to be temporally coherent in order for the video codec to take advantage of motion compensation. It is not at all clear that modulation-like high frequency patterns will be well-represented through motion compensation. Second, standard video codecs \cite{han2021technical,mukherjee2013latest,sullivan2004h,sze2014high, tudor1995mpeg,zhang2019recent}
 are substantially more complicated than image codecs, involving not only motion estimation and compensation, but also mode selection, loop filtering, and many other aspects. In this paper, we address these issues, while showing results for sandwiched video compression that parallel the earlier results for sandwiched image compression, including multispectral (color video compressed by grayscale video codecs) and super-resolution (high resolution video compressed by standard resolution video codecs) applications.  Moreover, we get gains over standard codecs with networks on the order of 100K parameters 
 (pre and post combined), a reduction in complexity by two orders of magnitude over state-of-the-art neural video compressors.  Few works prior to ours sandwich a standard video codec between neural processors. Those that do, e.g. \cite{Andreopoulos22,EusebioAP20,QiuLD21}, do so in such a way that the pre- and post-processors may be used independently --- thus do not take full advantage of the communication available between pre- and post-processors. Our contributions can be summarized as:

(1) We propose a sandwiched video compression framework that provides more efficient training and inference due to the simplicity of the neural networks but still enjoys the flexibility of neural networks in learning non-linear transforms.
     
(2) We propose a differentiable video codec proxy that can be used effectively in the training loop.
     
(3) We demonstrate that the sandwiched HEVC obtains 6.5 dB improvement over the standard HEVC 4:4:4 in low-resolution transport (8 dB improvement over 4:0:0 when coding color over gray-scale.) Moreover, when optimized for and tested with a perceptual similarity metric, Learned Perceptual Image Patch Similarity (LPIPS), we observe $\sim 30 \%$ improvements in rate at the same quality over the standard HEVC 4:4:4.

\vspace{-.4cm}
\section{Method}
\label{sec:method}
\subsection{Video Codec Proxy}

Our video codec proxy, shown in Figure~\ref{fig:diagram}, maps an input group of images (like a group of pictures, or GOP) to an output group of images, plus a bit rate for each image.  The details of our video codec proxy can be summarized as follows.

{\em Intra-Frame Coding}.  The video codec proxy simulates coding the first ($t=0$) frame of the group, or the I-frame, using the image codec proxy from \cite{guleryuz2021sandwiched,guleryuz2022sandwiched}.

{\em Motion Compensation}.  The video codec proxy simulates predicting each subsequent ($t>0$) frame of the group, or P-frame, by motion-compensating the previous frame, whether the previous frame is an I-frame or a P-frame.  Motion compensation is performed using a ``ground truth'' dense motion flow field obtained by running a state-of-the-art optical flow estimator, UFlow \cite{lodha1996uflow}, between the original source images $I_i(t)$ and $I_i(t-1)$.  The video proxy simply {\em applies} this ground truth motion flow to the previous reconstructed bottleneck image $\hat L_i(t-1)$ to obtain an inter-frame prediction $\tilde L_i(t)$ for bottleneck image $L_i(t)$.  This simulates motion compensation in a standard codec operating on the bottleneck images.  Note that our motion compensation proxy does not actually depend on $L_i(t)$, so although it is a warping, it is a linear map from $\hat L_i(t-1)$ to $\tilde L_i(t)$, with a constant Jacobian.  This makes optimization much easier than if we had tried to use a differentiable function of both $\hat L_i(t-1)$ and $L_i(t)$, like UFlow itself, that {\em finds} as well as {\em applies} a warping from $\hat L_i(t-1)$ to $L_i(t)$.  Such functions have notoriously fluctuating Jacobians that make training difficult.

{\em Prediction Mode Selection}.  Our video codec proxy also simulates Inter/Intra prediction mode decisions. This ensures better handling of temporally occluded/uncovered regions in video. First, Intra-prediction is simulated by rudimentarily compressing the current-frame and low-pass filtering it. This simulates filtering, albeit not the usual directional filtering, to predict each block from its neighboring blocks.  Initial rudimentary compression ensures that the Intra-prediction proxy is not unduly preferred at very low rates. For each block, the Intra prediction (from spatial filtering) is compared to the Inter prediction (from motion compensation), and the one closest to the input block determines the mode of the prediction.

{\em Residual Coding}.  The predicted image, comprising a combination of Intra- and Inter-predicted blocks, is subtracted from the bottleneck image, to form a prediction residual.  The residual image is then 
compressed using the image codec proxy of \cite{guleryuz2021sandwiched,guleryuz2022sandwiched}.  The compressed residual is added back to the prediction to obtain a ``pre-filtered'' reconstruction of the bottleneck image $\hat L_i(t)$.

{\em Loop Filtering}.  The ``pre-filtered'' reconstruction is then filtered by a loop filter to obtain the final reconstructed bottleneck image $\hat L_i(t)$.  The loop filter is implemented with a small U-Net((8);(8, 8)) \cite{UNet} that processes one channel at a time and is trained once for four rate points on natural video using only the video codec proxy with rate-distortion ($\ell_2$) training loss in order to mimic common loop filters. The resulting tandem of filters are kept fixed for all of our simulations, i.e., once independently trained, the loop filters for four rate points are not further trained.

{\em Rate Proxy}. Similar to \cite{guleryuz2021sandwiched,guleryuz2022sandwiched}, the differentiable rate for each frame is obtained as $    R(E_i(j)) = a\sum_{k,l}\log( 1+|e_l^{(k)}|/\Delta ),$ where $e_l^{(k)}$ is the $l$th coefficient of the $k$th block of DCT coefficients, and $a$ is chosen such that $R(E_i(j))$ is the actual rate at which JPEG codes $E_i(j)$ with uniform stepsize $\Delta$.
An alternative proxy is given in \cite{SaidSP22}.

\begin{figure}[t]
    \centering
    \includegraphics[width=\linewidth]{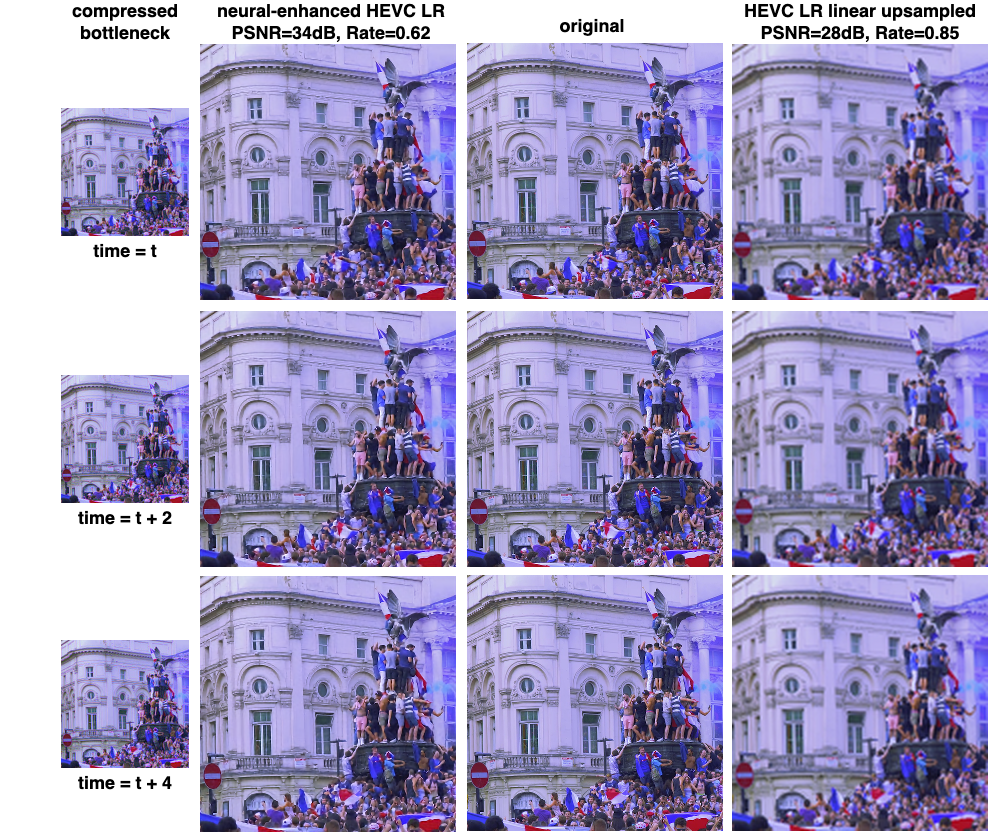}
    \caption{Sandwich of HEVC LR: Frames from compressed bottlenecks, reconstructions by sandwich, original source videos, and 
    HEVC LR.
    Note the substantially improved definition on the crowd and the building. While not shown here the sandwich system also obtains significant improvements over neural-post-processed HEVC LR thanks to the information embedded by the neural pre-processor. }
    \label{fig:subjective_LRHR}
\end{figure}

\subsection{Pre- and Post-Processors}

The pre- and post-processors are implemented as a $1\times1$ conv-network (a multilayer perceptron acting pixelwise) in parallel with a 
U-Net.  The intuition is that the $1\times1$ conv-network learns the appropriate color space conversion and tone mapping, while U-Net adds or removes spatial modulation patterns, or watermarks, which serve as the neural codes. We report results with U-Nets of two different complexities. As U-Net encoder/decoder filters, the high complexity network uses U-Net((32, 64, 128, 256); (512, 256, 128, 64, 32)) ($\sim 8$M parameters). Our low complexity, {\em slim} network uses U-Net((32); (32, 32)) ($\sim 57$ thousand parameters), with a $\sim \boldsymbol{ 99\%}$ reduction in parameters.

\subsection{Loss Function}

As mentioned in Section~\ref{sec:introduction}, the pre- and post-processors, and the loop filter in the video codec proxy are trained to minimize the total rate-distortion Lagrangian $\sum_{i,t}D_i(t)+\lambda R_i(t)$, where $D_i(t)$ and $R_i(t)$ are the distortion and rate of frame $t$ in clip $i$. The rate term in particular serves to encourage the pre- and post-processors to produce temporally consistent neural codes, since neural codes that move according to our ``ground truth'' motion flow field will be well predicted, leading to smaller residuals and rate terms.  We do indeed observe that the resulting neural codes seem to have some degree of temporal consistency, as shown in Figure~\ref{fig:subjective_YUV}. Note that the overall mapping from the input images through the pre-processor, video codec proxy, post-processor, and loss function is differentiable.

\vspace{-.25cm}
\section{Experiments}
\label{sec:experiments}
\subsection{Setup and Dataset Specifications}

\begin{figure}[h]
    \centering
    \includegraphics[width=\linewidth]{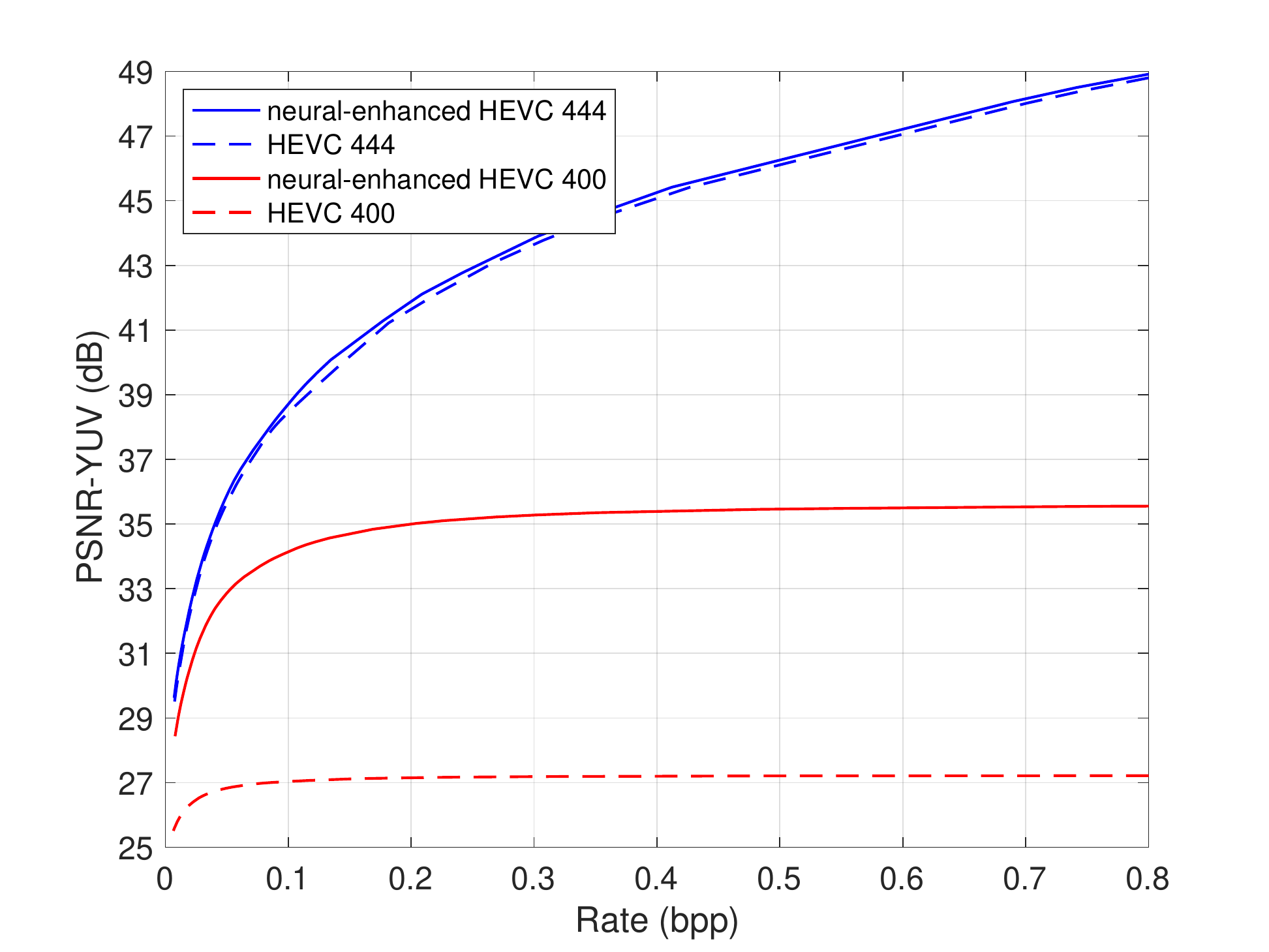}
    \caption{RD performance of the YUV 4:0:0 sandwich.}
    \label{fig:YUV400_comp}
\end{figure}

We generate a video dataset that consists of 10-frame clips of videos from the AV2 Common Test Conditions \cite{AV2CTC} and their associated motion flows, calculated using UFlow \cite{lodha1996uflow}. We use a batch size of 8, i.e., 8 video clips in each batch. Each clip is processed during the dataset generation step such that it has 10 frames of size 256x256, selected from video of fps 
20-40. HEVC is implemented using x265 (IPP.., single reference frame, rdoq and loop filter on.). We compare the sandwich model with the standard codec, HEVC, under three settings: (1) YUV 4:0:0 format with single-channel (grayscale) bottlenecks in Figure~\ref{fig:YUV400_comp}, (2) YUV 4:4:4 format in low-resolution (LR) bottlenecks in Figure~\ref{fig:LRHR_comp}, and (3) YUV 4:4:4 format using the Learned Perceptual Image Patch Similarity (LPIPS) -- a common perceptual similarity metric in the literature \cite{zhang2018unreasonable, elpips}. In each setting, the model is trained for 1000 epochs, with a learning rate 1e-4, and tested on 120 test video clips. {\em All RD plots are over the entire test set.}
We report results in terms of YUV PSNR when using the $\ell_2$ norm and ``LPIPS (RGB) PSNR'' when using LPIPS. Because LPIPS is intended for RGB, for the latter the final decoded YUV video is converted into RGB and then LPIPS is computed. In order to report results on an approximately similar scale, we derived a fixed linear scaler for LPIPS so that for image vectors $x, y$,
\begin{eqnarray}
s LPIPS(x, y) \sim ||x-y||^2, \mbox{if } ||x-y||^2 < \tau
\end{eqnarray}
where $\tau$ is a small threshold and $s$ is the LPIPS linear scaler. The LPIPS loss of a clip is the average of the LPIPS losses over 10 frames.

\subsection{Results}
\textbf{YUV 4:0:0.} Figures~\ref{fig:subjective_YUV} and~\ref{fig:YUV400_comp} show the subjective and quantitative results with sandwiched HEVC YUV 4:0:0 codec. As can be seen from the bottleneck frames in Figure~\ref{fig:subjective_YUV}, sandwich model is able to preserve the temporal consistency between the frames, which enables the standard codec HEVC to take advantage of the modulation patterns to achieve a better rate-distortion point. Quantitatively, Figure~\ref{fig:YUV400_comp} shows the substantial improvements (by 8 dB) of the sandwich model over 
HEVC YUV400. For reference HEVC 444 and its neural-enhanced version are also included. The latter  obtains compression-wise meaningful but relatively minor $5\%$ improvements in rate over the former. 

\begin{figure}[h]
    \centering
    \includegraphics[width=\linewidth]{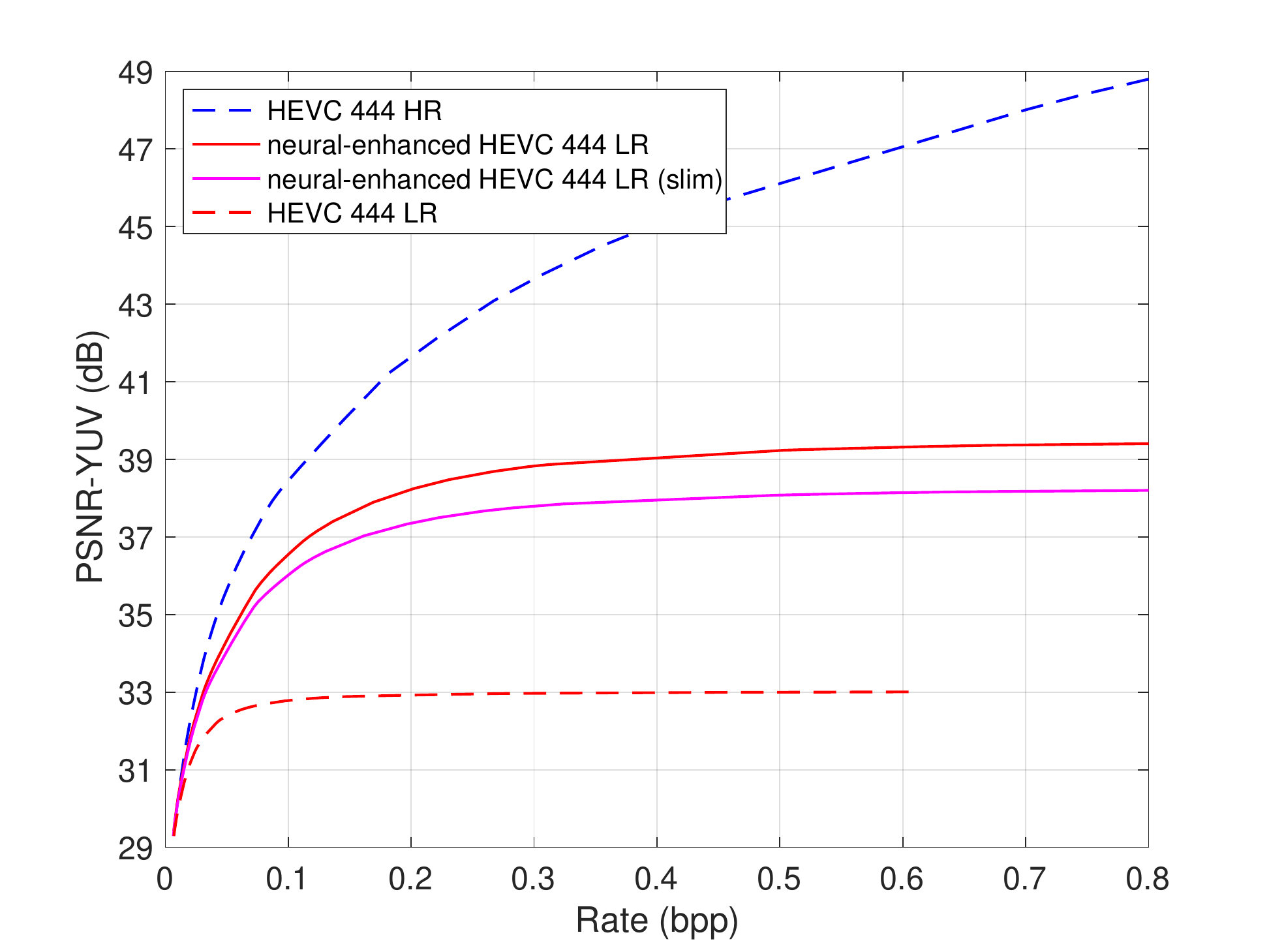}
    \caption{RD performance of YUV 4:4:4 low-resolution (LR) sandwich. (HEVC 444 HR stands for HEVC YUV 4:4:4 high-resolution.)}
    \label{fig:LRHR_comp}
\end{figure}

\textbf{Low-Resolution (LR).} 
Figures~\ref{fig:subjective_LRHR} and~\ref{fig:LRHR_comp} show the subjective and quantitative results with sandwiched HEVC YUV 4:4:4 LR codec. The standard LR codec transports LR video that is a linearly downsampled (bicubic) version of the original. Decoded video is linearly upsampled  (lanczos3). In the case of the sandwich, the exact same system is used but with generated bottlenecks. The 6.5 dB improvement of the sandwich model with HEVC YUV 4:4:4 LR  over the standard HEVC YUV 4:4:4 LR in Figure~\ref{fig:LRHR_comp} is also visible in the subjective comparison provided in Figure~\ref{fig:subjective_LRHR} which demonstrates the substantially detailed output of the sandwich in places where the standard HEVC LR produces blurred results.

\begin{figure}[t]
    \centering
    \includegraphics[width=\linewidth]{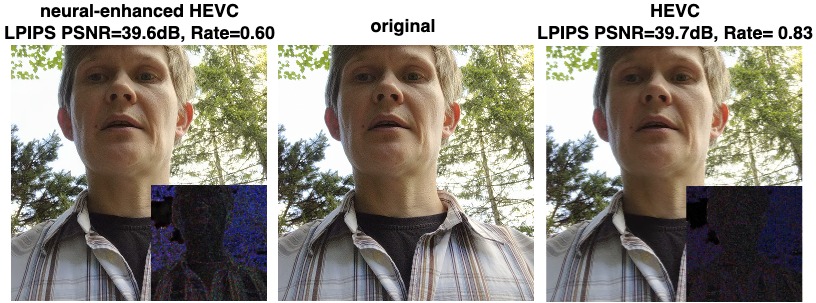}
    \caption{HEVC with LPIPS: Inter-frames from reconstructions by sandwich, original source videos, and reconstructions from HEVC (errors magnified 10x in the bottom-right). Note the rate reduction in the sandwich result. We were not able to observe perceptually meaningful differences despite larger absolute errors.}
    \label{fig:subjective_LPIPS}
\end{figure}
\begin{figure}[h]
    \centering
    \includegraphics[width=\linewidth]{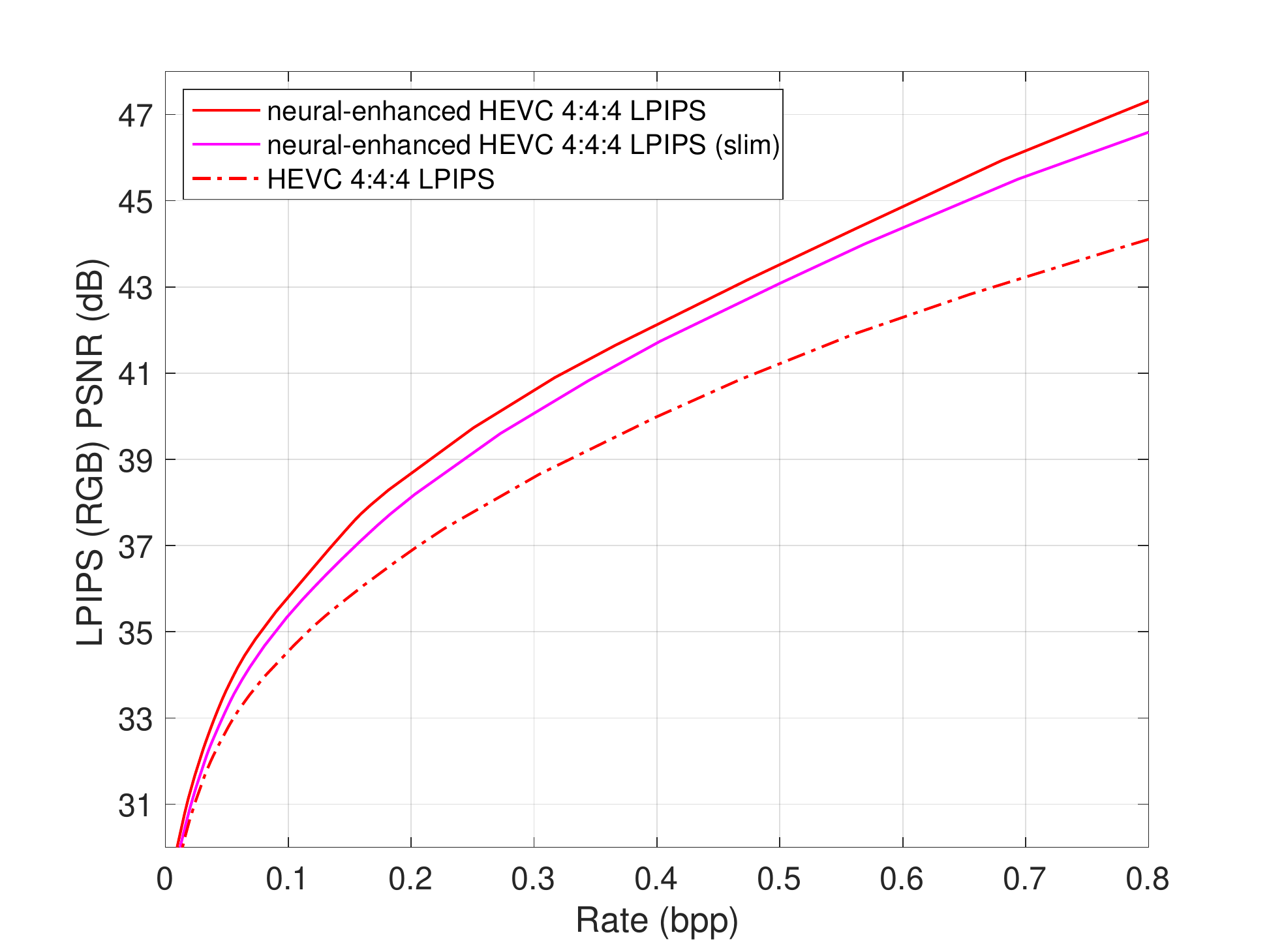}
    \caption{RD performance of the HEVC YUV 4:4:4 sandwich trained and tested with LPIPS. Neural-enhanced HEVC YUV 4:4:4 performs $\sim 30 \%$ better than the standard HEVC over a broad range of rates.}
    \label{fig:LPIPS_comp}
\end{figure}

\textbf{Perceptual Similarity Metric - LPIPS.} Finally, we train sandwich HEVC YUV 4:4:4 with LPIPS using a TensorFlow~\cite{tensorflow2015-whitepaper} implementation of \cite{zhang2018unreasonable} and compare it with the standard HEVC YUV 4:4:4 under LPIPS. The results are provided in Figures~\ref{fig:subjective_LPIPS} and~\ref{fig:LPIPS_comp}. Notice that the sandwich model provides $\sim 30 \%$ improvements in rate over the standard HEVC across a broad range.
Viewing the decoded video, we observe consistent reductions in rate obtained by the sandwich system with visually imperceptible differences (e.g., Figure~\ref{fig:subjective_LPIPS}.)
Concerns about the networks over-optimizing or ``hacking'' LPIPS are further alleviated by noting that because a clip typically depicts the same scene under slight geometric and color transformations, the utilized clip-averaged LPIPS loss  is expected to be robust for the scene--akin to the robustified ensemble LPIPS loss of \cite{elpips}.

\vspace{-.2cm}
\section{Conclusion}
\label{sec:conclusion}

In this work, we extend the sandwich architecture to video compression by carefully designing a video codec proxy and training it with neural pre- and post-processors in an end-to-end fashion. The proposed sandwich model outperforms the standard codec HEVC under various settings, including YUV 4:0:0 and YUV 4:4:4 LR formats; and under $\ell_2$ and LPIPS distortion, with gains of 8 dB (4:0:0), 6.5 dB (4:4:4 LR), and $\sim 30\%$ improvements in rate (LPIPS), respectively. We show that slim, light-weight networks with $57$K parameters can be used to closely approximate these results. The sandwich system can not only achieve a rate-distortion performance that is substantially superior to the standard video codec in these scenarios but it can do so without compromising computational efficiency through the use of light-weight networks. Our results clearly demonstrate that the sandwich system can re-purpose a standard codec to compression scenarios outside its immediate scope of design, from seamlessly increasing its resolution to optimizing it for a leading perceptual quality metric, all with significant improvements.

\bibliographystyle{abbrv} 
\bibliography{refs}
 \end{document}